\newcommand{\be}{\begin{equation} }
\newcommand{\ee}{\end{equation} }
\newcommand{\bea}{\begin{eqnarray} }
\newcommand{\eea}{\end{eqnarray} }

\newcommand{\muSR}{$\mu$-SR\xspace}
\newcommand{\PHCC}{(C$_4$H$_{12}$N$_2$)Cu$_2$Cl$_6$\xspace}
\newcommand{\dPHCC}{(C$_4$D$_{12}$N$_2$)Cu$_2$Cl$_6$\xspace}
\newcommand{\TCCL}{TlCuCl$_3$\xspace}
\newcommand{\KCCL}{KCuCl$_3$\xspace}

\documentclass[reprint,aps,prb,twocolumn]{revtex4-1}
\usepackage{amsmath,amssymb,bm}
\usepackage{graphicx}
\usepackage{color}
\usepackage{xspace}

\begin{document}


\title{High-pressure Raman study of the quantum magnet \PHCC.}


\author{S. Bettler}
\email{sbettler@phys.ethz.ch}
\author{G. Simutis}
\altaffiliation{Present address: Laboratory for Muon Spin Spectroscopy, Paul Scherrer Institut, 5232 Villigen PSI, Switzerland.}
\author{G. Perren}
\author{D. Blosser}
\author{S. Gvasaliya}
\author{A. Zheludev}
\email{zhelud@ethz.ch}
\homepage{http://www.neutron.ethz.ch/}
\affiliation{Laboratory for Solid State Physics, ETH Z\"urich, 8093 Z\"urich, Switzerland.}

\date{\today}

\begin{abstract}
Magnetic and lattice excitations in the quantum antiferromagnet \PHCC (PHCC) are studied across two pressure-induced phase transition at $P_c=4.3$~kbar and $P_1=13.4$~kbar using Raman spectroscopy. It is confirmed that neither transition is a result of a structural transformation. Magnetic  scattering is detected. It shows a pronounced pressure dependence and undergoes substantial changes at both transitions. The results are in clear contradiction with previous neutron studies, which detected only minor changes of the magnon spectrum at $P_1$. A number of phonons show anomalous frequency shifts at low temperatures. This effect is pressure dependent and for two of the observed phonons dramatically reverses sign at around $P_1$. The anomalous behavior is attributed to strong magnetoelastic coupling in PHCC.
\end{abstract}

\pacs{}

\maketitle

\section{\label{Introduction}Introduction}
 Quantum magnets are models of choice for studying the complex physics of quantum phase transitions.\cite{Sachdev2008} In such materials, quantum criticality can often be induced by an external magnetic field, as for example in the so-called Bose-Einstein condensation (BEC) of magnons \cite{Batyev1984,Giamarchi1999,Giamarchi2008} Another way to potentially drive these systems towards criticality is by applying external pressure and thereby tweaking the strength of magnetic interactions. In quantum paramagnets, i.e., spin systems with a quantum disordered spin-singlet ground state and a gap in the magnetic excitation spectrum, this type of perturbation may in rare cases lead to
a complete softening of the spin gap. The result is a unique magnetic soft-mode quantum phase transition to a long range-ordered state that breaks the $\mathrm{SO(3)}$ symmetry of the underlying Heisenberg Hamiltonian.\cite{Sachdev2008} The dynamical critical exponent of such a transition is $z=1$, in contrast to $z=2$ in field-induced BEC .

Only a handful of real materials undergo pressure-induced magnetic quantum phase transitions of the described  type. To date, only three examples are known: \TCCL \cite{Tanaka2003,Ruegg2004}, \KCCL \cite{Goto2006} and \PHCC (PHCC). In the latter, pressure-induced ordering was first detected by \muSR at $P_c=4.3$~kbar.\cite{Thede2014} Inelastic neutron scattering studies confirmed the transition to be of a soft-mode nature.\cite{Hong2010-2,Perren2015} A second transition was detected in PHCC by \muSR at a higher pressure of $P_1\sim 13.4$~kbar.\cite{Thede2014} It clearly corresponds to a rather drastic change in the magnetic structure, and is possibly a Lifshitz (commensurate to incommensurate) transition. Surprisingly though, inelastic neutron scattering experiments \cite{Perren2015} did not detect any drastic changes in the excitation spectrum of PHCC upon crossing $P_1$. Also unknown is to what extent the crystal lattice and phonons are involved, if at all, in either of the two transitions. The very limited high pressure neutron diffraction data available\cite{Perren2015} are a by-product of dedicated spectroscopy studies,  and are not conclusive on whether the transitions are accompanied by any crystallographic transformations. All these  issues we address in the present study by means of high pressure Raman spectroscopy.

\begin{figure}[h!]
\centering
\includegraphics[width=\columnwidth]{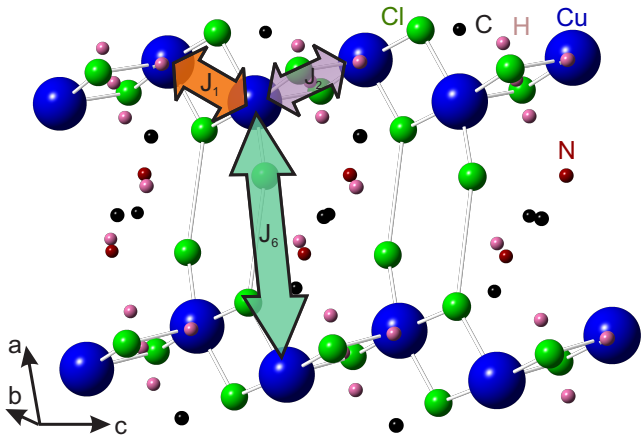}
\caption{A sketch of the crystal structure of PHCC, showing the three principal magnetic exchange interactions according to Ref.~\onlinecite{Stone2001}.}
\label{Raman_structure}
\end{figure}

PHCC is an extensively studied and well understood quantum paramagnet with a spin gap $\Delta=1.0$~meV.\cite{Stone2001,Stone2006,Stone2006-Nature} The crystal structure is triclinic, space group P$\bar{1}$, with lattice parameters $a=7.984(4)$, $b=7.054(4)$, $c=6.104(3)$~\AA , $\alpha=111.23(8)^\circ$, $\beta=99.95(9)^\circ$, and $\gamma =81.26(7)^\circ$. \cite{battaglia1988} The magnon dispersion relation is known from neutron experiments,\cite{Stone2001} with the magnon band extending between $\Delta$ and 2.7~meV. The topology of magnetic interactions between the $S=1/2$ Cu$^{2+}$ ions is believed to be rather complicated. The exchange constants are {\em not known}, but some conclusions can be drawn from an analysis of magnetic bond energies.~\cite{Stone2001} According to the latter, the three most important interactions roughly lie in the $(a,c)$ crystallographic plane and are schematically shown in  Fig.\ref{Raman_structure}. To some degree the system can be viewed as composed of antiferromagnetic dimers defined by the nearest neighbor exchange $J_1$. However, bond energies associated with a frustrated second-nearest-neighbor coupling $J_2$ and the sixth-nearest-neighbor antiferromagnetic bond $J_6$ are only slightly, if at all, smaller. There is evidence that it is the bond $J_1$ that is most affected by the application of hydrostatic pressure.\cite{Perren2015} With increasing pressure this coupling becomes progressively weaker, more than halving at 9~kbar.

\section{Experimental details}

In the present study, we utilized fully deuterated PHCC single crystals from the same growth batch as in Ref.~\onlinecite{Huevonen2012}. The samples were typically 0.2~$\times$~0.1~$\times$~0.1~mm$^3$ in size. In all cases the studied surface contained a well-defined $c$-axis, as verified by single crystal X-ray diffraction. Note that the strongest dimer bond is almost parallel to that direction.
The Raman experiments were performed using a Trivista 557 triple-grating spectrometer and a liquid-nitrogen-cooled charge-coupled device (CCD) detector. A 660-nm solid state laser at a low power of 3~mW was used in order to limit sample heating.  To achieve high stray light rejection as well as resolution, 1800/1800/1800~l/mm gratings were used.
Sample environment was a He-flow cryostat with a base temperature of 2.6~K. Measurements were carried out in a backscattering geometry using a focusing microscope. The data were taken in several acquisitions which allowed increasing the signal to noise ratio and excluding spurious events primarily due to cosmic  muons. For each spectrum, the total counting time was of the order of 1~h. The polarization of incident light could be rotated using a $\lambda/2$ plate. The polarization of scattered light was analyzed with a linear polarizer used in combination with a $\lambda/4$ plate.

\begin{figure}
\centering
\includegraphics[scale=1]{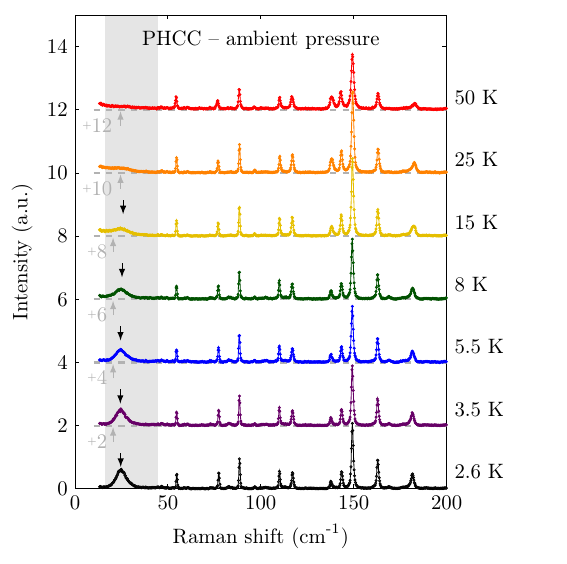}
\caption{Measured temperature dependence of the Stokes Raman scattering spectra in PHCC in $\bar{\mathrm{Z}}\text{(XX)Z}$ polarization at ambient pressure. Individual spectra are shown with incremental offsets for clarity. The arrows indicate the peak position of magnetic scattering. }
\label{Raman_temperature_magnetic}
\end{figure}
We define the polarization directions as X=$( 001)$, Y=$( 1\bar{1}0)$ with the momentum along Z=$( 110)$.
 With respect to the dominant magnetic couplings, these axes form following angles: $J_1$: X--23$^\circ$, Y--67$^\circ$;
$J_2$: X--24$^\circ$, Y--67$^\circ$; and $J_6$: X--90$^\circ$, Y--45$^\circ$, respectively. A configuration in which both incident and outgoing polarization are aligned along the crystallographic $c$-axis is thus denoted in the Porto notation as $\bar{\mathrm{Z}}\text{(XX)Z}$. $\bar{\mathrm{Z}}\text{(YY)Z}$ indicates that both polarizations are transverse to the direction of the $c$-axis and $\bar{\mathrm{Z}}\text{(XY)Z}$ denotes a cross-polarization geometry.

High-pressure measurements were done in a  miniature opposing plates diamond anvil cell with a BeCu gasket and backing plates based on the design of Ref. \onlinecite{Sterer90}. The cell utilized diamonds with a culet diameter of 1~mm and a gasket hole with diameter 0.6~mm. The pressure-transmitting medium was Daphne Oil 7373. The maximum pressure attained at base temperature was 18.2~kbar. The applied pressure was measured in situ with a precision of 0.1~kbar, using the shift of the ruby R1 fluorescence line.\cite{Feng2010}  The pressure homogeneity across the cell was about 1~kbar,  as estimated from measurements on several ruby spheres at different positions within the cell.

\section{Results}

\subsection{Ambient pressure}
\subsubsection{Magnetic scattering}
\label{ambientmag}
At elevated temperatures, the measured Raman spectra are dominated by phonons. Below about 25~K, a broad continuum of excitations develops at low energies as shown in Fig.~\ref{Raman_temperature_magnetic}. This continuum is well separated from any phonons, which allows for a clean observation of its pressure-dependence without the necessity to subtract any lattice contributions.
At $T=2.6$~K the continuum is peaked at 24.6 cm$^{-1}$ and has a full width at half maximum of 6~cm$^{-1}$ as extracted from empirical Lorentzian fits. It is almost symmetric with a very slight tail extending towards higher energies. Its intensity increases with decreasing temperature, although its position remains largely unchanged. This is a strong indication that the low-energy feature is due to magnetic scattering.\cite{Lemmens2000,Lemmens2003,Simutis2016Gvasaliya}

\begin{figure}
\centering
\includegraphics[scale=1]{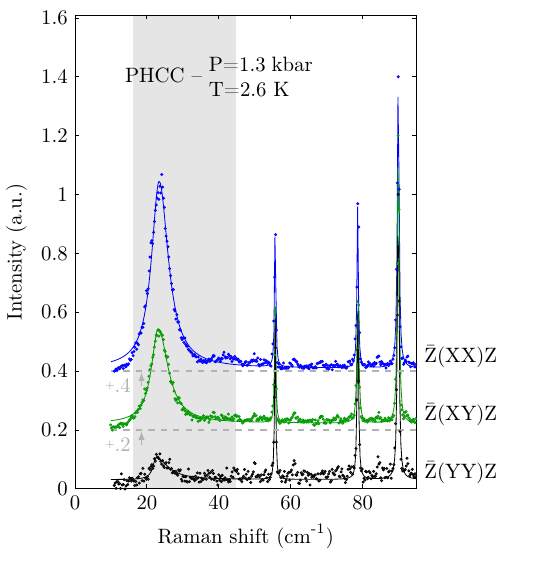}
\caption{Symbols: Measured polarization dependence of the Stokes Raman scattering spectra in PHCC at base temperature at 1.3~kbar. The shaded area is as in Fig.~\ref{Raman_temperature_magnetic}. Lines are empirical Lorentzian fits. Individual spectra are shown with incremental offsets for clarity.}
\label{Raman_polarization_magnetic}
\end{figure}

Another confirmation comes from analyzing its energy range. Magnetic Raman scattering in quantum paramagnets is expected to originate primarily from two-magnon processes. \cite{Lemmens2003,FleuryLoudon68} A broad continuum is indicative of a substantial magnon bandwidth. For gapped quantum antiferromagnets such a continuum is confined in the energy range between $2\Delta$ and twice the maximum magnon energy. For PHCC, using the single magnon dispersion known from neutron studies,\cite{Stone2001} we can etimate the domain of this 2-magnon continuum to be between 16~cm$^{-1}$ and 45~cm$^{-1}$. This range is indicated by the shading in Figs.~~\ref{Raman_temperature_magnetic} and \ref{Raman_polarization_magnetic}, and indeed coincides with the domain of the observed continuum.

The magnetic nature of the continuum is also consistent with the observed polarization dependence of scattering.
As shown in Fig.~\ref{Raman_polarization_magnetic}, the  shape and position of the continuum is the same in all experimental configurations. However, its intensity is strongest in the $\bar{\mathrm{Z}}\text{(XX)Z}$ geometry and much suppressed in the $\bar{\mathrm{Z}}\text{(YY)Z}$ geometry. Thus, the signal is strongest for polarization along the two strongest magnetic bonds, similar to two-magnon scattering in other known dimer systems.\cite{Lemmens2003,Choi2005}


%
%
\subsubsection{Anomalous phonons}
\begin{figure*}
\centering
\includegraphics[width=\textwidth]{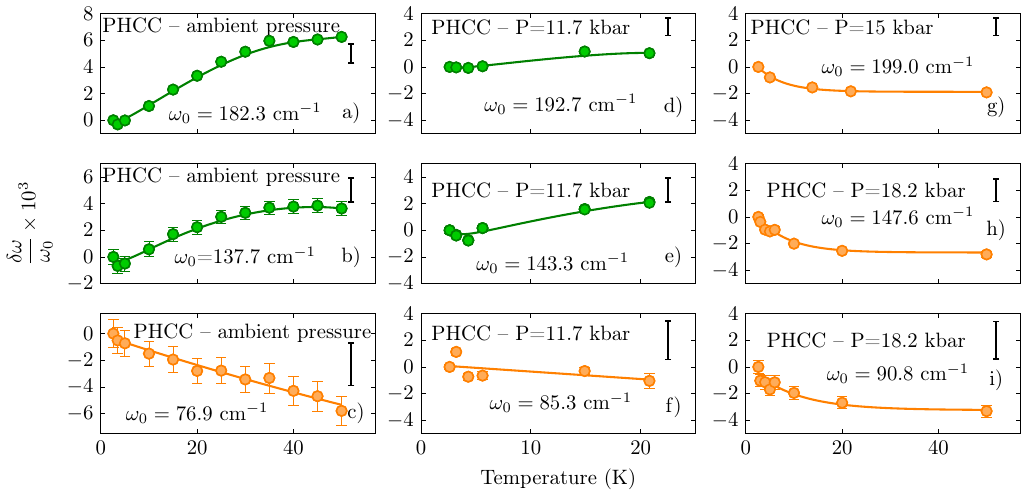}
\caption{Relative frequency shifts of three representative anomalous phonons in PHCC vs. temperature at ambient pressure, just below the second transition and above the transition, as deduced from Lorentzian fits to the measured spectra.
The detector bin sizes are shown in the upper right corners of each plot. Note that since each phonon peak is several detector bins wide, postulating its Lorentzian shape allows to determine the peak position with an accuracy greater than the bin size. Lines are guides to the eye. }
\label{Anomalous_phonon_Tdep}
\end{figure*}

In the energy range 10-205~$\mathrm{cm}^{-1}$ that was studied in detail in our experiments, 11 strong phonon peaks were observed. Since the crystal is triclinic with space group P$\bar{1}$, point group $C_i$, all Raman active phonons have trivial A$_g$ symmetry.
From the measured spectra, the phonon frequencies were obtained in Lorentzian fits.
As shown in Fig.~\ref{Anomalous_phonon_Tdep}, several of them exhibit anomalous temperature dependencies. Rather significant frequency shifts occur at {\em very low temperatures}, below 20~K. In this regime, two of the observed modes actually {\em harden} with increasing temperature.

In a bid to identify the eigenvectors of the anomalous phonons we performed density functional theory (DFT) {\em ab initio calculations} using the Quantum Espresso software package \cite{QUANTUM_ESPRESSO_2009}. The non-local rVV10 functional \cite{Sabatini2013} was used to incorporate Van der Waals and hydrogen bonding effects which are certainly important in this organomatallic compound. For all calculations the projector augmented wave method was used. Pseudopotentials for the density functional suggested in Ref.~\onlinecite{Vydrov2010} were generated using input parameters from the PSLibrary project (pslibrary.0.3.1).\cite{DALCORSO2014} In all calculations the kinetic energy cutoff was 120~Ry and the charge density cutoff
600~Ry. Brillouin zone integration was performed using a 3x3x3 k-point grid. 

For the 16 lowest-energy Raman-active modes the observed and calculated frequencies are listed in incremental order in the first two columns of Table~1. Additionally, measured and calculated phonon frequencies of a protonated sample are shown in the last two columms. The first 15 modes listed correspond to displacement patterns involving both the [Cu$_2$Cl$_6$]$^{2-}$ bitetrahedra as well as the organic ions. By contrast, the last mode listed in Table~1~--~as well as all higher-energy modes~--~correspond to internal molecular modes of the piperazinium ion. Consequently this last mode shows a much larger shift in frequency upon H/D exchange.

A conclusive result of our DFT simulations is that the 10- to 205-$\mathrm{cm}^{-1}$ low-energy lines that were studied in detail, all correspond to vibrations of the [Cu$_2$Cl$_6$]$^{2-}$ bitetrahedra with admixtures of low-energy vibrations of the piperazinium ions. This explains why almost all low-frequency vibrations are subject to substantial H/D isotope effect. At the same time, it greatly complicates the vibrational landscape and makes the frequencies very sensitive to the difficult to account for Van-der-Waals interactions. Therefore, it appears impossible to unambiguously match the observed anomalous phonon lines to particular calculated ones. Few studies of [Cu$_2$Cl$_6$]$^{2-}$ vibrations in other materials have been reported to date. In a recent work,\cite{OUESLATI201310} the authors also met considerable difficulties in assigning the low-frequency peaks to particular modes, as in our case.
%
\begin{center}
\begin{table}
\caption{Comparison of calculated and measured phonon energies for \dPHCC and \PHCC for the lowest vibrational modes in ascending order.}
\resizebox{\columnwidth}{!} {
\begin{tabular}{c c c c c}
\hline
\hline
\multicolumn{2}{c}{\dPHCC}& &\multicolumn{2}{c}{\PHCC}\\
\cline{1-2} \cline{4-5}
Expt.(cm$^{-1}$)&DFT(cm$^{-1}$)&\hspace*{10pt}&Expt.(cm$^{-1}$)&DFT(cm$^{-1}$)\\
\hline
54.4	&	49	&&	57.0	&	52	\\
76.9	&	74	&&	77.1	&	75	\\
88.3	&	76	&&	91.5	&	78	\\
96.4	&	85	&&	103.9	&	88	\\
109.8	&	93	&&	112.5	&	99	\\
116.7	&	105	&&	124.6	&	113	\\
137.7	&	117	&&	139.4	&	120	\\
143.2	&	126	&&	149.5	&	130	\\
149.0	&	137	&&	152.8	&	140	\\
162.7	&	148	&&	164.8	&	151	\\
182.3	&	160	&&	184.7	&	160	\\
206.1	&	181	&&	207.0	&	184	\\
278.2	&	258	&&	278.3	&	258	\\
290(2)	&	275	&&	290(2)	&	275	\\
305.3	&	302	&&	304.7	&	303	\\
319.9	&	305	&&	406.1	&	387	\\
\hline
\hline
\end{tabular}}
\label{calculation_comparison}
\end{table}

\begin{figure}
\includegraphics[scale=0.89]{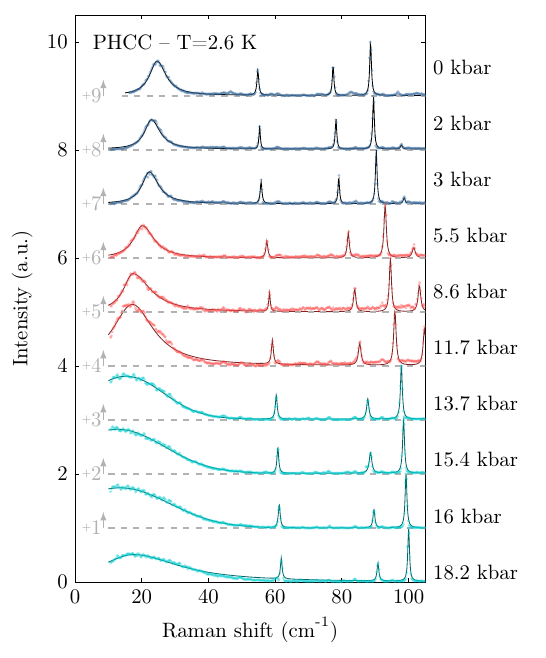}
\caption{Stokes Raman scattering spectra measured in PHCC at $T = 2.6$~K in $\bar{\mathrm{Z}}\text{(XX)Z}$ polarization, normalized to the peak intensity of the phonon with $\omega_0$=88.5~cm$^{-1}$. Individual spectra are shown with incremental offsets for clarity. The solid lines are guides for the eye obtained in empirical Lorentzian (0--12~kbar and 18.2~kbar) or Gaussian (13--16~kbar) fits to the continuum contribution.}
\label{Raman_pressure_magnetic}
\end{figure}
\end{center}
\begin{figure}
\includegraphics[scale=1]{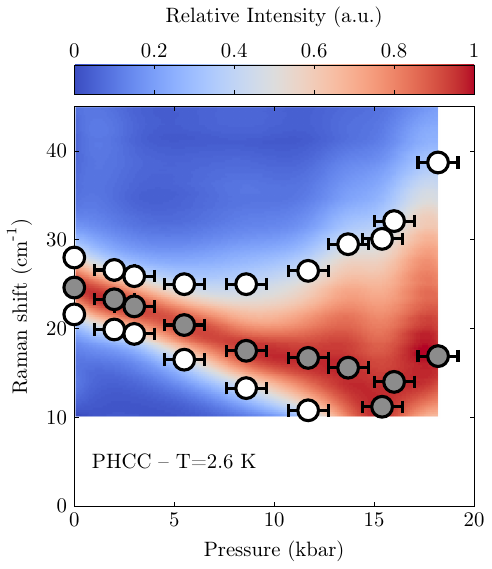}
\caption{Peak position and half-maximum frequencies of continuum magnetic scattering as extracted from Stokes spectra at $T = 2.6$~K in $\bar{\mathrm{Z}}\text{(XX)Z}$ polarization. }
\label{half_max}
\end{figure}
\subsection{Results under applied pressure}

\subsubsection{Magnetic excitations}
As could be expected, magnetic Raman excitations in PHCC are strongly affected by hydrostatic pressure. As shown in Fig. \ref{Raman_pressure_magnetic}, we observe a shift of the magnetic continuum to lower energies already at rather modest pressures. This is fully consistent with the previously observed softening of the spin gap.\cite{Hong2010-2} Below $P_c\sim 4$~kbar, the spectral function retains its symmetric shape with a slight tail towards higher energies. Throughout this pressure range the continuum width and shape remain constant, while it shifts to lower energies as a whole (Fig.~\ref{Raman_pressure_magnetic}).
In the intermediate-pressure range between $P_c$  and $P_1\sim 13$~kbar, we additionally observe a progressive change in the shape of the scattering. Whereas in the low-pressure range the asymmetry of the magnetic peak was subtle, the high-energy tail becomes considerably more pronounced at higher pressure. Increasing the pressure  above $\sim P_1$ results in a further significant change in the shape of the spectrum. The magnetic scattering peak becomes rounded and overdamped, with no clear maximum. Increasing the pressure further the spectrum continues to be broadened and develops a long high-energy tail.

Unlike in simple spin systems such as spin ladders,\cite{Brenig2001, Schmidt2001}  in a material with magnetic interactions as complex as they are in PHCC a quantitative analysis of the Raman spectrum does not appear feasible. To date there exist no model calculations of even the single-magnon dispersion in PHCC, let alone the Raman spectrum. As previously mentioned, even describing the ground state in terms of $J_1$-dimers is only approximate, since interdimer coupling is almost as strong. Instead, we chose a model-independent way to quantify the measured spectra. In Figure \ref{half_max} we plot the positions of the maximum of magnetic scattering, and frequencies where the intensity reaches half of its maximum value. The color map in the background represents the measured Raman intensity relative to its maximum value at each pressure.

\subsubsection{Phonons}

\begin{figure}
\centering
\includegraphics[scale=1]{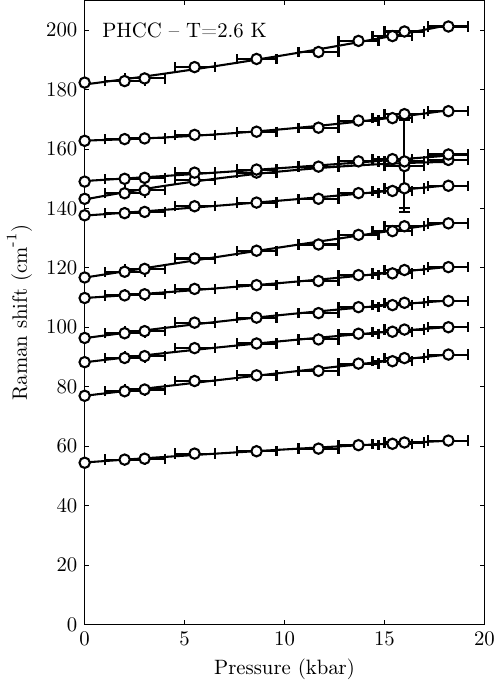}
\caption{Pressure dependence of the phonon frequencies measured in PHCC at $T=2.6$~K. Lines are guides to the eye.}
\label{Phonons_Pdep}
\end{figure}

Tracking the measured phonon frequencies vs. pressure confirms that the two transitions at $P_c$ and $P_1$ are not associated with any crystallographic transformations. Indeed, as shown in Fig.~\ref{Phonons_Pdep}, all  phonons observed at base temperature show a smooth and almost linear pressure dependence (hardening). None of disappearing, splitting, or an appearance of new phonon lines was detected up to 18.2~kbar applied pressure.

Despite the monotonic pressure dependence at a fixed low temperature, most of the observed phonons show a remarkable pressure-induced evolution of their temperature dependencies.  As shown in Fig.~\ref{Anomalous_phonon_Tdep}, the magnitude of the low-temperature frequency shifts is highly pressure dependent. For the two phonons that harden with increasing temperature at ambient pressure, even the {\em sign} of the temperature shift reverses under pressure. This reversal occurs between 11.7~ and 15~kbar, and thus seemingly coincides with $P_1$.

\section{Discussion}
As argued above, the magnetic scattering continuum can be ascribed to scattering from two-magnon processes that span between twice the energy gap and twice the maximum of magnon dispersion.  Within this interpretation, the observed evolution of the spectrum below $P_c$ implies a progressive softening of the gap without much change in the magnon bandwidth, in agreement with the neutron results. The only other comparable high pressure Raman study of a quantum paramagnet that we are aware of is that on KCuCl$_3$.\cite{Kuroe2012} The much broader resolution, stronger elastic line and lower signal-to-noise ratio of the spectra reported in that work make a direct comparison to our data difficult.

What in our optical measurements on PHCC is clearly in contradiction with the previous neutron experiments of Ref.~\onlinecite{Perren2015} are the observed steady increase of the magnon bandwidth above about 10~kbar, as deduced from the behavior of the upper continuum half-height frequency, and the clear changes of the shape of the scattering across the transition at $P_1$. In contrast, inelastic neutron scattering detected only insignificant changes in spin dynamics between 9~kbar and 18~kbar. Even those were only revealed in a quantitative fit to the measured neutron spectra, particularly to neutron intensities. The most likely reason for this discrepancy is an incorrect pressure calibration in the neutron study. As only became apparent through recent experience with the pressure cell used in the neutron experiment,\cite{Podleznyak} the actual pressure can drop by as much as a third, compared to nominal, upon cooling the cell down to base temperature. This effect was not taken into account in Ref.~\onlinecite{Perren2015}. Its magnitude for that particular experiment can not be assessed without additional \emph{in situ} pressure measurements using the same sample as in the original study. It appears likely though, that the nominally 18~kbar neutron data set actually corresponds to a pressure just below $P_1$.  Note that in the present optical study the pressure is known reliably, as it is measured \emph{in situ} at the experimental temperature. We thus confirm that the transition at $P_1$ is, indeed, accompanied by a substantial change in spin excitations and a rapid increase of the excitation bandwidth.

We can expect a direct connection between the evolution of the magnetic ground state and the behavior of anomalous phonons. First, note that the anomalous phonon behavior occurs below about 20~K. In most conventional materials, phonon frequencies remain constant at such low temperatures. The only energy scale in PHCC that would be consistent with the low temperature range is that of magnetic exchange interactions. Below 20~K is precisely where short-range spin correlations set in. For the strongest bonds, the typical exchange energy in PHCC is about 1~meV,\cite{Stone2001} which is larger but comparable to the observed anomalous frequency shifts (about 0.15~meV for the $\omega_0=181.3$~cm$^{-1}$ mode). With a high degree of certainty the anomalous behaviour can therefore be attributed to magnetoelastic coupling.  The pressure dependence of anomalous frequency shifts signifies a strong pressure dependence of the local spin correlations in the system. A phase transition to a qualitatively different magnetic structure that we expect to occur at $P_1$ will thus have a particularly strong effect. We can speculate that  it may be responsible for the observed sign reversals of two anomalous frequency shifts. A more concrete discussion of the microscopic mechanism would require an unambiguous identification of the  the eigenvectors for the anomalous modes, which is presently lacking.

\section{Conclusion}
Neither of the pressure-induced magnetic transitions in PHCC involve crystallographic transformations. The transition at $P_c$ is associated with a progressive softening of the magnon gap with little change of bandwidth. Contrary to previous indications, the transition at $P_1$ leads to a substantial reorganization of the spin excitation spectrum. The corresponding change of the magnetic ground state has a dramatic impact on the temperature dependencies of certain phonon modes. It appears that a further neutron scattering study of the high pressure phase may bring further insight into its origin.

\acknowledgments
We thank Dan Huevonen for his involvement in the early stages of the present project. We are grateful to Leonid Dubrovinsky for sharing his expertise on diamond anvil cells. We thank G\'abor Cs\'ucs, Victoria Elisabeth Blair, Peter John Beltramo, and Jan Vermant for technical assistance. Numerical simulations were performed on the Euler cluster (ETH Zurich).

\bibliography{PHCC_PRESSURE_31-8-2017_2}

\end{document}